\documentclass[12pt]{article}
\begin{document}

\title{Dark matter models with suppressed dark matter nuclei elastic  cross section} 
\author{N.V.Krasnikov 
\\Institute for Nuclear Research RAS
\\ Moscow 117312, Russia
\\and
\\ JINR Dubna, Russia
}
%\date{October,1997}
\maketitle
\begin{abstract}

 \vspace{0.2cm}

We propose two generalizations of the dark photon model which predict the suppressed
 elastic dark matter nuclei cross section  in comparison
with the corresponding prediction of the  dark photon model. In the first model the main difference
from dark photon model is that the mixing parameter $\epsilon$ is nonlocal formfactor
$\epsilon(q^2)= \frac{q^2}{\Lambda^2}V(\frac{q^2}{\Lambda^2}) $ depending on the square of the momentum transfer $q^2$.
Here $V(\frac{q^2}{\Lambda^2})$ is an entire function of the growth $\rho \geq \frac{1}{2}$ and $\Lambda$ is nonlocal scale.
  In this model
our world and dark world are described by renormalizable field theories while the
communication between them is performed by nonlocal interaction. The second model
is renormalizable model where 
besides dark photon field $A'$   additional   vector boson $Z'$   interacts with
$B - L$ current. The communication between our world  and dark world is performed due to nonzero kinetic mixing
between $Z'$ and $A'$ fields.  
The predictions of the models for the search for dark matter at the
accelerators  don't contain additional suppression factors.

\end{abstract}

\newpage

%\noindent
%PACS: 44.25.$+$f; 44.90.$+$c
%PACS No.: ~$12.60.-i, 12.60.Cn, 13.20.Cz$

%\label{sec:intro}

 \section{Introduction}
At present the most striking evidence in favour of new physics beyond
the Standard Model (SM) is the observation of the Dark Matter (DM) (as a review see
for example: \cite{ Kolb:1990vq, Rubakov:2017xzr  }).
A lot  of DM models exist \cite{Lindner}(as a review see for example 
\cite{Lindner, Gninenko:2020hbd, Gninenko:2021})
\footnote{In
  particular, there are Light Dark Matter (LDM) models 
  \cite{Boehm:2003hm, Gninenko:2020hbd, Gninenko:2021})
  with a
  mass of LDM particles $O(1)~MeV \leq m_{\chi} \leq O(1)~GeV$.}.
The standard assumption    \cite{ Kolb:1990vq, Rubakov:2017xzr  }        is that in the hot early
Universe the DM particles were in the equilibrium with the observed
particles is often used.
During the Universe expansion the temperature
decreases and at some point the thermal decoupling of the DM starts.
Namely, at some temperature the annihilation cross section of DM
particles 
$$
DM ~particles \rightarrow SM ~particles
$$
becomes too small to perform the equilibrium of DM particles with the SM
particles and the DM particles decouple. The experimental data are in favour of the scenario with cold relic for
which the freeze-out temperature is much lower the  DM particle mass, i.e. the nonrelativistic DM particles decouple.
The value of the DM annihilation cross section at the decoupling epoch determines the value  of the current DM particles.
The observed value of  the DM density fraction $\frac{\rho_{DM}}{\rho_c} \approx 0.23 $ allows to estimate the DM
annihilation cross section  and hence to estimate the DM  discovery potential.
There are at least three possibilities  to discover the  DM particles.
  At first it is possible to  use the  LHC experiments  ATLAS, CMS, LHCb 
    or fixed target experiments like NA64 and BELLE. The second way is the use of the
   elastic DM nucleon(electron) scattering in underground experiments \cite{PARTICLEDATA}.
   The third  possibility consists in the astrophysical   detection of the DM annihilation into the
   SM particles \cite{PARTICLEDATA}.
   At present the underground experiments
  give bounds on the elactic DM nucleon cross section at the
  level   $O(10^{-43} - 10^{-47})~cm^2$   \cite{PARTICLEDATA}  that strongly restricts a lot
  of existing DM models \cite{Lindner}.   Therefore it is interesting to consider the  DM models which allow
  to escape the strong bounds  from underground  experiments and don't have any additional suppression factors for
  the accelerator experiments. 

  In this paper we propose two generalizations of the dark photon model \cite{DARKPHOTON}  with suppressed DM nucleon(electron)
  elastic cross section.   In our models the elastic tree level
    DM nucleon(electron) cross section has suppression factor\footnote{Here $v$ is the DM velocity
    in speed of light units.} $O(v^4) = O(10^{-12})$  in comparison
    with the corresponding prediction for the  dark photon model. At one loop level
    the suppression factor for the cross section is around $O(10^{-6})$.
   In the first model the main difference
from dark photon model is that the mixing parameter $\epsilon$ is nonlocal formfactor
$\epsilon(\frac{q^2}{\Lambda^2})= \frac{q^2}{\Lambda^2}V(\frac{q^2}{\Lambda^2}) $ depending on the square of the momentum transfer $q^2$.
Here $V(\frac{q^2}{\Lambda^2})$ is an entire function of the growth $\rho \geq \frac{1}{2}$ and $\Lambda$ is nonlocal
scale.   In the proposed  model the SM world and the DM  world are described by renormalizable field theories while the
communication between them is performed by nonlocal interaction. The second model is renormalizable model
  where besides dark photon $A'$  additional  $Z'$ vector boson interacts with
$B - L$ current of the SM. The interaction of the SM world and the  DM world is performed only due to nonzero kinetic mixing of
$A'$ and $Z'$ fields.   
%Both proposed  models predict the p-wave annihilation of dark matter that allows to 
%escape Planck bound \cite{PLANCK} on LDM.  
  The predictions of the models for the search for dark matter at the
accelerators and in astrophysics  don't contain additional suppression factors.

The predictions of the models for the search for DM  at the
accelerators and in astrophysics don't contain additional suppression factors.

  The paper is organized as follows. In the next section we describe  nonlocal generalization of the dark photon model.
    In the third section we consider renormalizable model with
    additional $U(1)$ $Z'$ vector boson interacting with dark photon field $A'$.
    Section 4 contains concluding remarks. Appendix contains  the main formulae used for
  the DM density calculations.

%\label{sec:The LDM phenomenology with dark photon messenger}
\section{Nonlocal generalization of dark photon model}

In dark photon model \cite{DARKPHOTON}
%\cite{Holdom:1985ag,1982xi}
  the additional light vector boson $A'$ interacts with the gauge
$SU_c(3) \otimes SU_L(2) \otimes U(1)$ fields of the SM  due to nonzero mixing with the $U(1)$ SM
gauge field. The Lagrangian of the model is represented in the form
\begin{equation}
L = L_{SM} + L_{SM,dark} + L_{dark} \,,
\label{1}
\end{equation}
where   $L_{SM}$ is the SM Lagrangian and  
\begin{equation}
L_{SM,dark} = -\frac{\epsilon}{2\cos\theta_w}B^{\mu\nu}F'_{\mu\nu} \,.
\label{2}
\end{equation}
Here $B^{\mu\nu} = \partial^{\mu}B^{\nu} - \partial^{\nu}B^{\nu}$,   
$F'_{\mu\nu} = \partial_{\mu}A'_{\nu} - \partial_{\nu}A'_{\mu}$, $\epsilon$ is the mixing parameter,  and  $L_{dark}$
is the DM   Lagrangian \footnote{The field  $B_{\mu}$ is the  $U(1)$ gauge field of the SM.}.
At present scalar, Dirac,  pseudo-Dirac and Majorana DM models are often considered. 
%For instance, for the scalar DM the Lagrangian  $L_{dark}$ has the form
%\begin{equation}
%L_{dark} = -\frac{1}{4}F'_{\mu\nu}F'^{\mu\nu}   
%+ (\partial_{\mu}\chi - ie_D A_{\mu}^{'} \chi)(\partial^{\mu}\chi - ie_D A^{'\mu} \chi)^* - m^2_{\chi}\chi^*\chi 
%%-\lambda_{\chi}(\chi^*\chi)^2
%+  \frac{m^2_{A'}}{2}A'_{\mu}A'^{\mu} \,,
%\label{3s}
%\end{equation}
%where $\chi$  is the charged scalar DM field.

In dark photon model with Dirac or scalar DM  the  electron DM elastic cross section has the form \cite{USA}
 \begin{equation}
   \sigma(DM ~+~electron \rightarrow DM ~+~electron) = \mu^2_{\chi e} \frac{16\pi \epsilon^2\alpha \alpha_D}{m^4_{A'}} \,,
  \label{elDMcross1}
 \end{equation}
 where $\mu_{\chi e} = \frac{m_{\chi}m_e}{m_{\chi} + m_e}$ and $\alpha_D = \frac{e^2_D}{4\pi} $.
% is the analog of electromagnetic coupling constant
% $\alpha = \frac{1}{137}$.
 The analogous formula is valid for nucleon.  For $m_{\chi} \approx 10^3~GeV$ experimental bounds on
 $ \sigma(DM ~+~nucleon \rightarrow DM ~+~nucleon)$ are at the level $10^{-9}~pb$ \cite{PARTICLEDATA} that restricts rather
 strongly dark photon
 mass,  namely $m_{A'}  \geq 3.5~TeV $ at $\alpha_D = 0.1$ and $\epsilon = 0.1$.

 %In dark photon model with the Majorana DM the elastic cross section (\ref{elDMcross1})
 %is suppressed by factor $k_M =\frac{2\mu^2_{\chi e}}{m^2_{\chi}} v^2$.

 In nonlocal  generalization of the dark photon model  we assume that both  our world and  dark world are described by
 local renormalizable field theories while the communication between our world and dark sector is performed by nonlocal interaction
 \cite{Efimov1, Efimov2, Efimov3, Krasnikov1}.
  We propose to use  nonlocal generalization for the mixing term
(\ref{2}), namely 
\begin{equation}
  L_{SM,dark}
  %= -\frac{\epsilon}{2\cos\theta_w}B^{\mu\nu}F'_{\mu\nu}
  \rightarrow
L_{SM,dark~nonlocal} = -\frac{1}{2\cos\theta_w}B^{\mu\nu}\epsilon(-\frac{\partial^{\mu}\partial_{\mu}}{\Lambda^2})  F'_{\mu\nu}   \,.
\label{2nonlocal}
\end{equation}
In nonlocal field theory the formfactor   $ \epsilon(\frac{q^2}{\Lambda^2})$ is an entire function on $q^2$ of the growth
$\rho \geq \frac{1}{2}$ \cite{Efimov1, Efimov2, Efimov3, Krasnikov1}.
   Moreover we require that nonlocal interaction  $ \epsilon(\frac{q^2}{\Lambda^2})$
has to dissapear in the limit of the infinite nonlocal scale $\Lambda$, i.e.
$ \epsilon(\frac{q^2}{\Lambda^2})   \rightarrow 0 $  at $\Lambda \rightarrow \infty$.
  In other worlds it means that  the communication  between our world and dark world switches off for
infinite nonlocal scale.
As a consequence we find that
$\epsilon(\frac{q^2}{\Lambda^2}) = \sum _{k=1}^{\infty}c_k(\frac{q^2}{\Lambda^2})^k$. As an example we shall use the formfactor
\begin{equation}
  \epsilon(\frac{q^2}{\Lambda^2}) =
  \frac{q^2}{\Lambda^2}\exp(-  \frac{(q^2)^2}{\Lambda^4} ) \,.
\label{nonlocform}
\end{equation}
The existence of exponential multiplier $\exp(-  \frac{(q^2)^2}{\Lambda^4})$ in the formfactor (\ref{nonlocform})
leads to  ultraviolet convergence of the corresponding Feynman diagrams.
The Feynman rules for the nonlocal dark photon model are the same as in original dark photon
model except the use of the formfactor $\epsilon(\frac{q^2}{\Lambda^2})$ instead of the constant $\epsilon$.
It should be stressed that  nonlocal formfactor $\epsilon(\frac{q^2}{\Lambda^2}) \rightarrow 0$ for $q^2 \rightarrow 0$ that is crusial
for the suppression of the elastic DM nucleon(electron) cross section.  
In nonlocal model with the formfactor (\ref{nonlocform}) vanishing at $q^2 \rightarrow 0$ we have the suppression factor
$k_{nl} = \frac{16v^4 \mu^4_{\chi N}}{3\Lambda^4} \sim O(10^{-12})$
%\footnote{For $m_{\chi} \gg m_N$
%  $k_{nl}  =O(10^{-12})\frac{m^4_N}{\Lambda^4}$.} 
for  elastic cross section  (\ref{elDMcross1}).
In nonlocal dark photon model the
production cross section of dark photon $A'$ is proportional to $\epsilon^2(\frac{m^2_{A'}}{\Lambda^2})$.
%In this paper we assume that in the early Universe the DM was in the equilibrium with the SM.
%as a consequence we have the equation for the determination of  the annihilation cross section.
For nonlocal dark photon model the formula for the annihilation cross section
$\sigma(DM ~+~ \bar{DM} \rightarrow SM ~particles)$
coincides with the corresponding formula for the standard dark photon model except
the replacement $\epsilon^2 \rightarrow \epsilon^2(\frac{4m^2_{\chi}}{\Lambda^2})$. For $q^2 \ll \Lambda^2$ and
$m_{A'} = k m_{\chi}$ we find that $\epsilon(\frac{m^2_{A'}}{\Lambda^2}) =
    \frac{k^2}{4}\epsilon(\frac{4m^2_{\chi}}{\Lambda^2})$.
%\begin{equation}
%  \epsilon(4 m^2_{\chi}) = \frac{4}{k^2}\epsilon(m^2_{A'}) \,.
%\end{equation}
%For often used reper case $m_{A'} = 3 m_{\chi}$
% In particular $ \epsilon(\frac{4 m^2_{\chi}}{\Lambda^2}) = \frac{4}{9}\epsilon^2(\frac{m^2_{A'}}{\Lambda^2})$
%for often used reper case   $m_{A'} = 3 m_{\chi}$. In dark photon  model
Using the formulae of the Appendix
we can  estimate the product $\alpha_D\epsilon^2(\frac{4m^2_{\chi}}{\Lambda^2})$ as a function of $m_{\chi}$ and $m_{A'}$.
%One can find that both the case of LDM and DM with $m_{\chi}$ can be realized for some mass parameters.

Consider at first the case of scalar LDM with $m^2_{A'} = 5 m^2_{\chi}      $ and  $\alpha_D = 0.1$.
For  dark photon  mass $m_{A'} \leq O(1)~GeV$ the NA64 \cite{NA64.1} and BABAR \cite{BABAR}
experiments give the most strongest bounds on $\epsilon({\frac{m^2_{A'}}{\Lambda^2}})$.
As a consequence of the assumed equilibrium of the LDM with the SM particles  at the early Universe one can find
    that
\begin{equation}
  \epsilon(\frac{m^2_{A'}}{\Lambda^2}) \sim 0.9 \cdot 10^{-6} (\frac{m_{A'}}{MeV}) \,.
  \label{nonlocest1}
\end{equation}
For pseudo- Dirac LDM with $m_{A'} = 3 m_{\chi}$ and $\alpha_D - 0.1$ we find that 
\begin{equation}
  \epsilon(\frac{m^2_{A'}}{\Lambda^2}) \sim 1.5 \cdot 10^{-6} (\frac{m_{A'}}{MeV}) \,.
  \label{nonlocest1a}
\end{equation}
The obtained values (\ref{nonlocest1}, \ref{nonlocest1a})  for  $\epsilon(\frac{m^2_{A'}}{\Lambda^2})$ don't  
contradict to experimental bounds
 \cite{PLANCK}, \cite{ NA64.1, BABAR}   at  $m_{A'} \leq 1~GeV$ . The predicted value of nonlocal scale
$\Lambda$
depends on dark photon  mass $m_{A'}$ and it is rather small, for instance $\Lambda \sim  10~GeV$
at $m_{A'} = 100~MeV$. 
%The  experimental bounds on the mixing parameter $\epsilon$ in standard dark photon model are
%valid for the mixing parameter $ \epsilon(m^2_{A'}) $ of the nonlocal dark photon model.

For the  mass region    $m_{\chi}  \sim O(1)~TeV$ consider as an example  fermion   dark matter
with $\alpha_D = 0.1$ and $m^2_{A'} = 5 m^2_{\chi}$.  The analog of the formula (\ref{nonlocest1})
reads
\begin{equation}
  \epsilon(\frac{m^2_{A'}}{\Lambda^2}) \sim 0.08  (\frac{m_{A'}}{TeV}) \,.
  \label{nonlocest2}
\end{equation}
The predicted value of nonlocal scale $\Lambda$ depends on the dark photon mass $m_{A'}$,  for
instance   $\Lambda = 7~TeV $  at   $m_{A'} = 2~TeV $.
Note that the obtained value
(\ref{nonlocest2}) for the mixing parameter does not contradict to the LEP1 data
since the mixing parameter  $\epsilon(\frac{q^2}{\Lambda^2})$ strongly depends on the mass scale. 
In the energy region of the $Z$-boson we have to use  $\epsilon(\frac{m^2_{Z}}{\Lambda^2}) =
\frac{m^2_Z}{m^2_{A'}}\epsilon (\frac{m^2_{Z}}{\Lambda^2}) $  (for $m_{A'} = 2~TeV$
$\epsilon(\frac{m^2_{Z}}{\Lambda^2}) \approx  0.2 \cdot 10^{-3}$) 
which  is suppressed by factor $\frac{m^2_Z}{m^2_{A'}}$ in comparison with  $\epsilon(\frac{m^2_{A'}}{\Lambda^2})$.
For often used mass relation $m_{A'} = 3 m_{\chi}$ and $\alpha_D = 0.1$ we find that  
$\epsilon(\frac{m^2_{A'}}{\Lambda^2}) \sim 0.54  (\frac{m_{A'}}{TeV})$ and $\Lambda = 1.4~TeV$ at
$m_{A'} = 1~TeV$.
%For instance, for the Dirac DM with $m_{A'} = 3 m_{\chi}$ and $\alpha_D = 0.1$ we
%obtain  that for  $m_{\chi} =1~GeV, 100~GeV, ~200~GeV, ~400~GeV, ~1000~GeV$
%$\epsilon(\frac{4m^2_{\chi}}{\Lambda^2}) = 0. 002,  ~0.04,~ 0.08, ~0.16, ~0.4  $
%and $\Lambda = 45~GeV,~ 1~TeV, ~1.4~TeV, ~2~TeV,  3.2~TeV $. For the
%Dirac DM with  $m^2_{A'} = 5 m^2_{\chi}$  and 
%    $\alpha_D = 0.4$ we find that for $m_{\chi} = 10~TeV$  $\epsilon(\frac{4m^2_{\chi}}{\Lambda^2}) =    0.4  $
%and $\Lambda = 32~TeV$.
%For nonlocal dark photon model with dark photon  mass $m_{A'} \leq 10~GeV$ the NA64 \cite{NA64.1} and BABAR \cite{BABAR}
%experiments give the most strongest bounds on $\epsilon{\frac{m^2_{A'}}{\Lambda^2}})$.

It should be stressed that in considered model we have huge suppression $O(10^{-12})$ only at tree level.
At one-loop level the elastic DM nucleon cross section
has suppression factor $O(\frac{1}{8\pi^2})\epsilon^2(\frac{m^2_{A'}}{\Lambda^2})\alpha\alpha_D$
  in comparison with tree level cross section (\ref{elDMcross1}).
%\begin{equation}
%\sigma(DM ~+~nucleon \rightarrow DM ~+~nucleon)
%\sim m^2_N \frac{ \epsilon^4(\frac{m^2_{A'}}{\Lambda^2})\alpha^2 \alpha_D^2}{\pi m^4_{A'}}
%\label{oneloop}
%\end{equation}
  For $\epsilon(\frac{m^2_{A'}}{\Lambda^2}) =0.1$ and  $\alpha_D = 0.1$ the suppression factor is $O(10^{-7})$.

 %one-loop cross section (\ref{oneloop})
%i%s suppressed by factor $O(10^{-7})$ in comparison to tree level cross section (\ref{elDMcross1}).

The CMS and ATLAS  bounds for nonlocal dark photon model with a mass $m_{A'} \geq O(200)~GeV$ coincide with the corresponding bounds
for  dark photon model \cite{PARTICLEDATA,ATLAS1} and they  are not very strong. The reason is that
%in contrast to the most
%of the $Z'$-models 
dark photon decays mainly into invisible modes  $A' \rightarrow \chi \bar{\chi}$
that makes its detection not very easy in contrast to
most $Z'$   models(for instance the $Z'$ model with $B-L$
current) which have visible decays into $e^+e^-$ or $\mu^+\mu^-$. Moreover in dark photon model the cross section of the dark photon 
production is suppressed by factor $ \epsilon^2(\frac{m^2_{A'}}{\Lambda^2}) $.

Within nonlocal approach it is possible to  use  nonlocal interaction between the SM fields and
the DM fields without dark photon. For instance,  nonlocal interaction of
the  ($vector \times vector$) type
%can play the role of messenger between the SM world and dark
% sector. Namely, let us consider  the nonlocal interaction
 \begin{equation}
   S_{nl, messenger} \supset  \int d^4xJ^{\mu}_{SM}(x)V(-\frac{\partial^{\mu}\partial_{\mu}}{\Lambda^2})J_{\mu,dark} \,
   \label{nonlocalint}
 \end{equation}
 can play the role of messenger between the SM world and dark
 sector.
 Here $J^{\mu}_{SM}$ is the current made up of  the SM fields, $J_{\mu,dark}$ is the current made up
 of the dark particles and $ V(-\frac{\partial^{\mu}\partial_{\mu}}{\Lambda^2}) $ is nonlocal formfactor.
% As an example we
% use  the $(B-L)$ current $J^{\mu}_{B-L} = \sum_{leptons}\bar{l}\gamma^{\mu}l -
% \frac{1}{3}\sum_{quarks}\bar{q}\gamma^{\mu}q  $   and nonlocal formfactor (\ref{nonlocform}).
%\begin{equation}
%  V(p^2) =  \frac{p^2}{\Lambda^4} \exp(- (\frac{p^2}{\Lambda^2})^2 +1) \,.
% \label{nonlocform}
%  \end{equation}
%  We take the  $J_{\mu, dark}$
%   equal to  $J_{\mu, dark}= \bar{\chi}_D\gamma_{\mu}\chi_D$ for Dirac DM and
%  $ J_{\mu, dark}= \frac{1}{2}\bar{\chi}_M\gamma_{\mu}\chi_M $ for Majorana DM.
 For the model (\ref{nonlocalint}) with the formfactor  (\ref{nonlocform})  the elastic
 tree level DM  nucleon(electron) cross section is suppressed by the same factor $O(v^4)$.

 \section{Renormalizable extension of the dark photon model  with additional vector $Z'$ boson}
 In dark photon model dark photon field $A'$ interacts with DM particles
 due to nonzero kinetic mixing between  the dark photon field
 $A'$ and the $U(1)$ gauge field $B$ of the SM model. Here
 we consider the extension of the dark
 photon model with  additional $U(1)$ gauge field $Z'$
 interacting with the SM fields. As the simplest possibility we consider the interaction of
 the $Z'$ with $(B-L)$ current \cite{Davidson, Marshak, Okada, Okada2}, namely
 \begin{equation}
   L \supset   g_{B-L}Z_{\mu}'
   (\sum_{leptons}\bar{l}\gamma^{\mu} l
   - \frac{1}{3} \sum_{quarks}
     \bar{q}\gamma^{\mu}q )
 \end{equation}
 We assume that  dark photon  $A'$ interacts with the DM matter in standard way.
 For instance, for the Dirac fermion DM the interaction is
 \begin{equation}
   L \supset e_D\bar{\chi}\gamma^{\mu}\chi A_{\mu}' \,.
 \end{equation}
 Note that in DM model with ($B-L$) $Z'$ vector boson the DM nucleon elastic cross section is \cite{Lindner}
 \begin{equation}
   \sigma(DM ~+~nucleon \rightarrow DM ~+~nucleon) = \mu^2_{\chi N} \frac{g^2_{B-L}g^2_{\chi}}{\pi m^4_{Z'}} \,,
  \label{elDMcross2}
 \end{equation}
 Here $g_{\chi}$ is the coupling constant of DM with $Z'$.
 From the experimental bound $ \sigma(DM ~+~nucleon \rightarrow DM ~+~nucleon) \leq 10^{-9}~pb $ \cite{PARTICLEDATA}
 and the formula  (\ref{elDMcross2}) for the elastic cross section we find  rather  strong bound
 $m_{Z'} >  1.8\cdot 10^{4} \sqrt{g_{B_L}g_{\chi}}      ~GeV $.
 In our model we assume that $Z'$ does not interact directly with the DM. The interaction of $Z'$
 with DM is performed due to nonzero kinetic mixing of $Z'$ with dark photon $A'$, namely
 \begin{equation}
   L \supset -\frac{\epsilon_{Z'A'}}{2}A^{\mu\nu}Z'_{\mu\nu} \,,
   \label{epsiloninter}
   \end{equation}
 where $A^{\mu\nu}  = \partial^{\mu}A^{\nu '} - \partial^{\nu} A^{\mu '}$,
 $ Z'_{\mu\nu} = \partial_{\mu}Z_{\nu }' -\partial_{\nu}Z_{\mu }'  $. 
 As a consequence of the  interaction (\ref{epsiloninter}) and nonzero  $Z'$ mass we find that
 the tree level amplitudes with the interaction of  $Z'$ and $A'$ bosons   contain the multiplier
 $    \frac{q^2}{q^2 - m^2_{Z'}} $, where $q$ is the momentum transfer. Note that in dark photon model the role of the
 $Z'$-boson plays   massless photon field $A$ and the multiplier is $\frac{q^2}{q^2}   = 1$. As a
 consequence of nonzero $Z'$ boson mass
 for $|q^2| \ll m^2_{Z'}$ we have the suppression  factor $\frac{q^2}{m^2_{Z'}}$ for tree level  amplitudes.
 As it was explained in the previous section the existence of the 
 factor $\frac{q^2}{m^2_{Z'}}$
   leads to the suppression factor $O(v^2) = O(10^{-12})$ for the tree level elastic DM nucleon(electron) cross section.
 For the model with additional $Z'$ boson   consider two mass regions for DM. For the case of the LDM with $O(1)~MeV \leq
 m_{\chi} \leq O(1)~GeV $ there are rather strong bounds on coupling constant $g_{B-L}$ for the $Z'$ boson, see (\cite{NA64.2})
 and \cite{Bilmis, Lindner2, CHARM2}.
 %We can estimate the value of the annihilation cross section
 % $\sigma(DM~ + ~DM \rightarrow  SM particles)$ for considered model from the requirement that at the ealy Universe
 %the DM was in the equilibrium with the SM mateer.
 Consider as an example the scalar  LDM. The annihilation cross section into electron positron pair
  in nonrelativistic approximation has the form
  \begin{equation}
    \sigma(\chi \bar{\chi} \rightarrow  e^+ e^-)v_{rel} = \frac{ g^2_{B-L}e^2_D \epsilon^2_{Z'A'} m^2_{\chi}v^2_{rel} }{6\pi}
    (\frac{4m^2_{\chi}}{(4m^2_{\chi} - m^2_{Z'})(m^2_{A'} - 4 m^2_{\chi})})^2
    \label{csB-L}
  \end{equation}
  In comparison with the $B-L$ LDM model we have additional factor
  $k_{ad} = \epsilon^2_{Z'A'} (\frac{4m^2_{\chi}}{m^2_{A'} - 4 m^2_{\chi}})^2$
  for the cross section  (\ref{csB-L}). For the particular case
  $m^2_{A'} = 3 m^2_{\chi}$  and  $  \epsilon_{Z'A'} = 0.25 $ the additional factor  $k_{ad} = 1$ and the predictions
  for $g_{B-L}$ for both models coincide.
  Also the predictions of the considered  model with $m^2_{A'} = 3 m^2_{\chi}$, $m^2_{Z'} = 5 m^2_{\chi}  $, $ \epsilon_{Z'A'} = 0.05 $
  coincide with the predictions of the  ($B-L$) model with $m_{Z'} = 3 m_{\chi}$
 % In general we  find that  for $k_{ad} =  O(1))$ the  model can reproduce the observed  value of the LDM density
 % without  contradiction  to the experimental bounds on light $Z'$ boson.

  For the  mass region  with $m_{\chi} = O(1)~TeV$ the model also does not contradict to  existing accelerator
  bounds for some parameters.
  Consider  the model with Dirac DM.
 % where $ k = 6.5 $ for  $m_{\chi} \gg m_{top}$.
  For $m_{\chi} = O(1)~TeV$   the equation for the determination of the DM density leads to
  \begin{equation}
    \frac{k g^2_{B-L}e^2_D\epsilon^2_{Z'A'}   m^2_{\chi} }{\pi}
    (\frac{4m^2_{\chi}}{(4m^2_{\chi} - m^2_{Z'})(4 m^2_{\chi} - m^2_{A'}   )})^2  \sim 5 \cdot 10^{-9}~GeV^{-2} \,,
    \label{densecmod}
  \end{equation}
   where $ k = 6.5 $ for  $m_{\chi} \gg m_{top}$. 
As a numerical example we use   $g_{B-L} = e _{D} =1$, $m_{Z'}   = 3 m_{\chi}$, $m^2_{A'} = 5m^2_{\chi}$. 
  As a consequence of the equation (\ref{densecmod}) we find that
$\frac{\epsilon_{Z'A'}}{m_{Z'}} \approx 2 \cdot 10^{-5}~GeV^{-1}$. 
% $m_{Z'} = 8.3~TeV$. 
%  In particular, for $g_{B-L} = e _{B-L} =1$, $m^2_{Z'}   = 8 m^2_{\chi}$, $m^2_{A'} = 5 m^2_{\chi}$
%  and $m_{\chi} = 3~TeV$ we
%  find that $\epsilon_{B-L} \sim 0.13$.  
%   For $g_{B_L} = e _{B_L} =1$, $m^2_{Z'}   = 5 m^2_{\chi}$, $m^2_{A'} = 3 m^2_{\chi}$
%  and $m_{\chi} = 40~TeV$ we find that 
%  $\epsilon_{B-L} \sim 0.4$. 
From the LEP bound $\frac{m_{Z'}}{g_{B-L}} > 7~TeV $ \cite{Alcarez, Langacker}\footnote{For $g_{B-L} =1 $  LEP bound $m_{Z'} > 7~TeV$
  is stronger the LHC bounds.}  we find that the mixing
  parameter $\epsilon_{Z'A'} \geq 0.14$.
  As a second  numerical example consider  $g_{B-L} = e _{D} =1$, $m^2_{Z'}   = 5 m_{\chi}^2$, $m^2_{A'} = 3m^2_{\chi}$.
  For this set of parameters we find that $\frac{\epsilon}{m_{Z'}} \approx 0.54 \cdot 10^{-5}~GeV^{-1}$ and
   as a consequence of the LEP bound    \cite{Alcarez, Langacker}  $\epsilon_{Z'A'} \geq 0.038$.
 % Note that the obtained value  $m_{Z'} = 8.7~TeV$  does not contradict to the LHC bounds
%nd to the LEP bound $\frac{m_{Z'}}{g_{B-L}} > 7~TeV $ \cite{Alcarez, Langacker}. As a consequence  we
%find  that
%$\epsilon_{B-L}  > 0.125$  for  $g_{B-L} = e _{D} =1$, $m_{Z'}   = 3 m_{\chi}$, $m^2_{A'} = 5m^2_{\chi}$.
%Note that in our estimates we   used  the $Z'$ boson mass values
%  which   don't
%  contradict to the  LHC bounds on the $Z'$ boson mass.
  %So we find that in the model with two additional heavy vector bosons
  %we have thermo dark matter\footnote{Thermo dark matter means the DM  is
  %  in equilibrium with the SM matter in the early Universe} and the
  %cross section of the elastic DM electron(nucleon) is suppressed by factor
  %v^4 \sim O(10^{-12}$ that makes the direct observation of DM hopeless task.
  %The considered model predicts the existence of two new  vector bosons. The first new vector boson $Z'$
  % interacts with quarks and leptons. Here as an example we considered the
  %$Z'$-boson interacting with the $(B-L)$ current. The second vector boson  $A'$ (dark photon)  interacts only with
  % DM. The $Z-A'$ mixing  leads to the interaction
  %between the DM and  the SM particles.
  %The main peculiarity of the model is the prediction of the $O(v^4) = O(10^{-12})$
  %suppression factor for the elastic DM nucleon(electron) cross section.
   As in previous example at one-loop level we have the suppression factor
   $(\frac{g_{\chi} g_{B-L}\epsilon_{Z'A'}}{8\pi^2})^2 \sim O(10^{-6})$ in comparison to  tree level
     cross section  \label{elDMcross2} and  the bound on $Z'$-boson mass is weaker by factor $\sim 30$
     the corresponding bound  in $(B-L)$ DM model.
     The  $Z'$-boson phenomenology in considered model is similar to  the
  phenomenology of the $Z'$-model without dark photon. All LHC and fixed target 
  bounds are valid for the model with additional dark photon $A'$.
   In contrast to the $Z'$-boson
  dark photon $A'$ decays mainly into invisible modes.
   As a consequence the  bounds on
  dark photon are much weaker the bounds on $Z'$-boson. 
 % Note thtt for considered model it
 % is possible to have LDM with masses between $O(1)~MeV$ and $O(1)~GeV$ like in the
 % model with dark photon without contadiction to
 % existing bounds on $g_{B-L}$ coupling constant for $Z'$ boson. 

  \section{Conclusions}

  In this paper we proposed two generalizations of the dark photon model which predict the
  suppression for the  elastic DM nucleon(electron) cross section 
   in comparison with the corresponding prediction of the dark photon model.
   In the first model the main difference
   from dark photon model is that the mixing parameter $\epsilon$ is nonlocal formfactor
   $\epsilon(q^2)= \frac{q^2}{\Lambda^2}V(\frac{q^2}{\Lambda^2}) $ depending on the square of the momentum transfer $q^2$.
   Here $V(\frac{q^2}{\Lambda^2})$ is an entire function of the growth $\rho \geq \frac{1}{2}$ and $\Lambda$ is
   nonlocal scale.
 In this model
our world and dark world are described by renormalizable field theories while the
communication between them is performed by nonlocal interaction. In the second model
besides dark photon field $A'$ we introduce additional  $Z'$ vector boson interacting with
$B - L$ current. The interaction between our world and dark world is performed
only due to nonzero kinetic mixing of the $A'$ and $Z'$ fields.
Both models allow the existence of the LDM with a mass $m_{\chi} = O(1)~GeV$
and the DM with the mass in TeV region. The predictions  for the search for dark matter at the
 accelerators and in astrophysics don't contain additional suppression factors.

  I am indebted to my colleagues from INR RAS for discussions.

\section{Appendix. The main formulae for the DM density calculations}
In the nonrelativistic approximation with $<\sigma v_{rel}> = \sigma_0 x^{-n}_f$
one can find that \cite{Kolb:1990vq, Gondolo} takes place for the  $x_f$ calculation:
\begin{equation}   
  \Omega_{DM}h^2 = 0.1 (\frac{(n+1)x_{f}^{n+1}}{g_{*s}/g_*^{1/2}}) \frac{0.876 \cdot10^{-9} GeV^{-2}}{\sigma_0}\,.
\end{equation}
Here $x_f = \frac{m_{\chi}}{T_d}$, $T_d$ is the decoupling temperature, $\Omega_{DM}$ is the DM density of
the Universe, h is the value of Hubble parameters in special units and
$<\sigma v_{rel}>$ is the  product of the DM annihilation cross section and the relative velocity of the DM particle .
The value of $n=0$ corresponds to
the s-wave annihilation and the $n = 1$ corresponds to the p-wave annihilation.
The following approximate formula \cite{Kolb:1990vq, Sherrer, Cline} takes place for the  $x_f$ calculation:
\begin{equation}
  x_f = c - (n + \frac{1}{2})\ln(c) \,,
\end{equation}
\begin{equation}
c =\ln(0.038(n+1)\frac{g}{g^{1/2}_{*,av}}M_{PL}m_{\chi}\sigma_0) \,.
\end{equation}
Here $M_{PL} = 1.22 \cdot 10^{19} GeV$ is the Planck mass,  $g^{1/2}_{*,av} = \frac{1}{T_d}\int^{T_d}_{0}(g_{*s}/g^{1/2}_*)dT$,
$g_{*s}$($g_*$)  is the number of relativistic degree of freedom for entropy(energy)
and  $g$ is the number of internal degree of fredom for  DM particle $\chi$.
If DM particle differs from DM antiparticle $\sigma_o = \frac{\sigma_{an}}{2}$ where
$\sigma_{an}$ is the annihilation cross section of the reaction $\chi + \bar{\chi} \rightarrow
~ ~SM ~particles $.
In our numerical estimates we take  $\Omega_{DM}h^2 = 0.12$. For s-wave annihilation cross section with $n=0$
\begin{equation}
  <\sigma v_{rel}> = 7.3 \cdot 10^{-10}GeV^{-2}\frac{1}{g^{1/2}_{*,av}}(\frac{m_{\chi}}{T_d})\,.
\end{equation}
%Here $g^{1/2}_{*,av} = \frac{1}{T_d}\int^{T_d}_{0}(g_{*s}/g^{1/2}_*)dT$.
%For $1~MeV \leq m_{chi} \leq 1~GeV$ we have $10 \leq \frac{m_{\chi}}{T_d} \leq 20$. For 
%$T_f \leq 100~MeV$ the effective value $g^{1/2}_{*,av} \approx  3.3$.
%In our numerical estimates we use 
%and  $\frac{m_{\chi}}{T_d} = 15$.
%We find
%\begin{equation}
%  <\sigma v_{rel}>  = 3.3 \cdot 10^{-9} ~GeV^{-2}
%\end{equation}
For dark photon model with Dirac DM particles the annihilatio cross section into
electron positron pair has the form
\begin{equation}
  \sigma_{an}(\chi \bar{\chi} \rightarrow e^{-}e^{+})v_{rel} = \frac{16 \pi \epsilon^2 \alpha \alpha_D m^2_{\chi}}
        {(m^2_{A^`} - 4 m^2_{\chi})^2} \,.
\end{equation}
Here $\alpha =\frac{e^2}{4\pi}= 1/137$, $m_{A`}$ is the dark photon mass,
$\alpha_D = \frac{e_D^2}{4\pi}$  and $\epsilon$ is the mixing parameter.
  For the p-wave annihilation in nonrelativistic approximation $  <\sigma v_{rel}>
  = <B v^2_{rel}> = 6B \cdot \frac{T_d}{m_{\chi}}$. In dark photon model with scalar DM particles
  the annihilation cross section is
\begin{equation}
  \sigma_{an}(\chi \bar{\chi} \rightarrow e^{-}e^{+})v_{rel} = \frac{8\pi \epsilon^2 \alpha \alpha_D
    v^2_{rel} m^2_{\chi}}
        {3(m^2_{A^`} - 4 m^2_{\chi})^2} \,.
\end{equation}


\newpage


\begin{thebibliography}{99}






%  \def\selectlanguageifdefined#1{
%\expandafter\ifx\csname date#1\endcsname\relax
%\else\selectlanguage{#1}\fi}
%\providecommand*{\href}[2]{{\small #2}}
%\providecommand*{\url}[1]{{\small #1}}
%\providecommand*{\BibUrl}[1]{\url{#1}}
%\providecommand{\BibAnnote}[1]{}
%\providecommand*{\BibEmph}[1]{\emph{#1}}
%\ProvideTextCommandDefault{\cyrdash}{\hbox to.8em{--\hss--}}
%\providecommand*{\BibDash}{\ifdim\lastskip>0pt\unskip\nobreak\hskip.2em\fi
%\cyrdash\hskip.2em\ignorespaces}

%\bibliographystyle{unsrt}
%\bibliographystyle{pepan}
%\bibliography{maikbibl}



\bibitem{Rubakov:2017xzr}
    Valery A.Rubakov,  and Dmitry Gorbunov, Introduction to the Theory of the Early Universe,
  World Scientific Pub. Co., Singapore, 2017.

\bibitem{Kolb:1990vq}
  Edward W.Kolb and Michael Turner,  The Early Universe, \\
  FERMILAB-BOOK-1990-01, 1990.

\bibitem{Lindner} G.Arcadi at al., arXiv:2403.15860.

\bibitem{Gninenko:2020hbd}
S.N.Gninenko, N.V.Krasnikov, V.A. Matveev, 
Phys.Part.Nucl. {\bf 51}, 829 (2020): arXiv:2003.07257~[hep-ph].

\bibitem{Gninenko:2021}
  S.N.Gninenko, N.V.Krasnikov, V.A.Matveev,  
  Usp.Fiz.Nauk  {\bf 191}, 1361 (2021). 
%\newblock arXiv:2003.07257~[hep-ph].


\bibitem{Boehm:2003hm}
 C.Boehm and P.Fayet,
%\href{"10.1016/j.nuclphysb.2004.01.015    }
 Nucl.Phys. {\bf B683}, 219 (2004); \\
 arXiv:2003.07257~[hep-ph].

\bibitem{PARTICLEDATA} As a review see for example: \\
  R.L.Workman et al.(Particle Data Group), Progr.Theor.Exp.Phys. \\
  {\bf C01}, 08(2022) and 2023 update.
\bibitem{PLANCK} P.A,R.Ade  et al.(Planck Collaboration) \\ 
 Astron.Astrophys. A {\bf 13}, 594(2016).  
\bibitem{DARKPHOTON}
 Bob Holdom,  Phys.Lett. {\bf B166},196 (1986).

\bibitem{USA} Jim Alexander et al., arXiv:1608.08632.

\bibitem{Efimov1} G.V.Efimov,Nucl.Phys. {\bf 74}, 657 (1965).

\bibitem{Efimov2} G.V.Efimov, Comm.Math.Phys. {\bf 5}, 4 (1967).

\bibitem{Efimov3} G.V.Efimov, Ann.Phys. {\bf 71}, 466 (1972).

\bibitem{Krasnikov1} As a recent review see for example: \\
  N.V.Krasnikov, Nuovo.Cim. {\bf C45}, 30 (2022).
\bibitem{ATLAS1} G.Aad et al.(ATLAS collaboration) arXiv:2404.15741.

\bibitem{NA64.1} Yu.M.Andreev et al.(NA64 Collaboration), Phys.Rev.Lett. {\bf 131} 16, 161801 (2023).

\bibitem{BABAR}J.P.Lees et al.(BABAR collaboration), Phys.Rev.Lett. {\bf 119}\\
  131804 (2017).

  \bibitem{Davidson} A.Davidson, Phys.Rev. {\bf D20}, 776 (1979).

 \bibitem{Marshak} R.E.Marshak and R.N.Mohapatra, Phys.Lett. {\bf B91}, 222 (1980).

\bibitem{Okada} N.Okada and O.Seta, Phys.Rev. {\bf D82}, 023507 (2010).

\bibitem{Okada2} N.Okada and Y.Orikasa, Phys.Rev. {\bf D85}, 115006 (2012).

\bibitem{NA64.2} Yu.M.Andreev et al.(NA64 Collaboration), Phys.Rev.Lett. {\bf 129} 16, 161801 (2023). 

\bibitem{CHARM2} P.Vilain et al.,Phys.Lett.{\bf B302},351 (1993);
    {\bf B335},246 (1994).

  \bibitem{Bilmis} S.Bilmis et al., Phys.Rev. {\bf D92}, 033009 (2015).
  
\bibitem{Lindner2} M.Lindner et al., Phys.Lett. {\bf B811}, 135972 (2020).

\bibitem{Alcarez} J.Alcarez et al., arXiv:hep-ex/0612034.

\bibitem{Langacker} P.Langacker, Rev.Mod.Phys. {\bf 81}, 1199(2009).

  
\bibitem{Gondolo} P.Gondolo and G.Gelmini, Nucl.Phys. {\bf B360}, 145 (1991).
  
\bibitem{Sherrer} R.J.Sherrer  and M.S.Turner, Phys.Rev. {\bf D33}, 1585 (1986).

  \bibitem{Cline} J.M.Cline et al., Phys.Rev. {\bf D88}, 055025 (2013).
  
    
\end{thebibliography}
\end{document}